\documentstyle[sprocl,epsfig]{article}

\def\Journal#1#2#3#4{{#1} {\bf #2}, #3 (#4)}

\def\NPB{{\em Nucl.~Phys.}~B}
\def\NPA{{\em Nucl.~Phys.}~A}

\def\PRL{\em Phys.~Rev.~Lett.}
\def\PRD{{\em Phys.~Rev.}~D}
\def\ZPC{{\em Z.~Phys.}~C}
\def\be{\begin{equation}}
\def\ee{\end{equation}}
\def\bea{\begin{eqnarray}}
\def\eea{\end{eqnarray}}

\newcommand{\ul}{\underline}
\newcommand{\tstrut}{\rule[-2.4mm]{0mm}{8mm}}
\newcommand{\pslash}{\kern 0.2 em p\kern -0.4em /}
\newcommand{\Pslash}{\kern 0.2 em P\kern -0.56em \raisebox{0.3ex}{/}}
\newcommand{\Sslash}{\kern 0.2 em S\kern -0.56em \raisebox{0.3ex}{/}}

\begin{document}

\title{
\begin{flushright}
\small
hep-ph/9707340\\
NIKHEF 97-031\\[4mm]
\end{flushright}
DISTRIBUTION AND FRAGMENTATION FUNCTIONS IN A SPECTATOR MODEL
\footnote{
Talk at the Conference on Perspectives in Hadronic Physics,
Trieste, 12-16 May 1997.}
}

\author{\underline{R.~Jakob}, P.J.~Mulders  and J.~Rodrigues}

\address{NIKHEF, P.O. Box 41882, NL-1009 DB Amsterdam, the Netherlands}

\maketitle
\abstracts{Quark distribution and spectator functions are estimated in a
diquark spectator model. The representation of the functions in terms of
non-local operators together with the rather simple model allow estimates for
the yet experimentally undetermined functions, like the polarized fragmentation
functions or higher twist functions.
}

More details about the method and the results reported here can be found in the
long write-up~\cite{jmr97} by the same authors.

\section{What are the objects of interest ?}

The basic tool in the description of hard processes is the (assumption of)
factorization into hard and soft physics. The hard parts, i.e., partonic
cross-sections are calculated using the rules of perturbative QCD, whereas the
soft parts have either to be taken from experiment or calculated, for instance
with the help of models. The idea is that the non-perturbative objects
containing information on the hadronic confinement are universal, i.e., they
occur in all hard processes in the same form.

In a field-theoretic description of hard processes the soft,
non-perturbative physics is contained in (connected) hadronic matrix elements of
non-local operators built from quark and gluon fields. The simplest, but most
important ones are quark-quark correlation functions.
 
\subsection{quark-quark correlation functions}

The {\em quark-quark correlation function} containing information
about how partons 
are distributed inside a hadron with momentum $P$ and spin $S$ in the initial 
state of the process can be written for each quark flavor as~\cite{sop77,jaf83}
\be
\label{eq:Phi}
\Phi_{ij}(p,P,S)=\sum_X\int\frac{d^4x}{(2\pi)^4}\;
e^{ip\cdot x}\;\langle P,S|{\overline\psi_j(0)}|X\rangle
\langle X|{\psi_i(x)}|P,S\rangle.
\ee
The information on how a quark, with a specific flavor, fragments into a
hadron with momentum $P_h$ and spin $S_h$ (plus anything else; denoted by $X$)
is encoded in the correlation function~\cite{col82}
\be
\Delta_{ij}(k,P_h,S_h)=\sum_X\int\frac{d^4x}{(2\pi)^4}\;
e^{ik\cdot x}
\langle 0|{\psi_i(x)}|P_h,S_h;X\rangle
\langle P_h,S_h;X|{\overline\psi_j(0)}|0\rangle .
\ee
To be specific we consider the case were the hadrons are protons.
\vspace{2mm}
\begin{center}
\includegraphics[width=4.5cm]{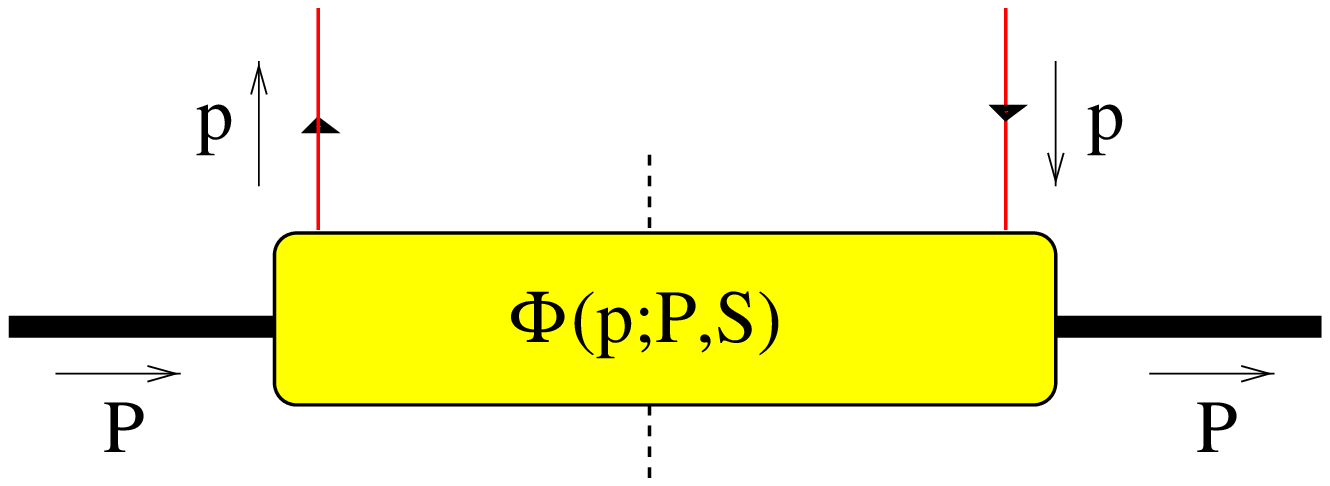}
\hspace{10mm}
\includegraphics[width=3cm]{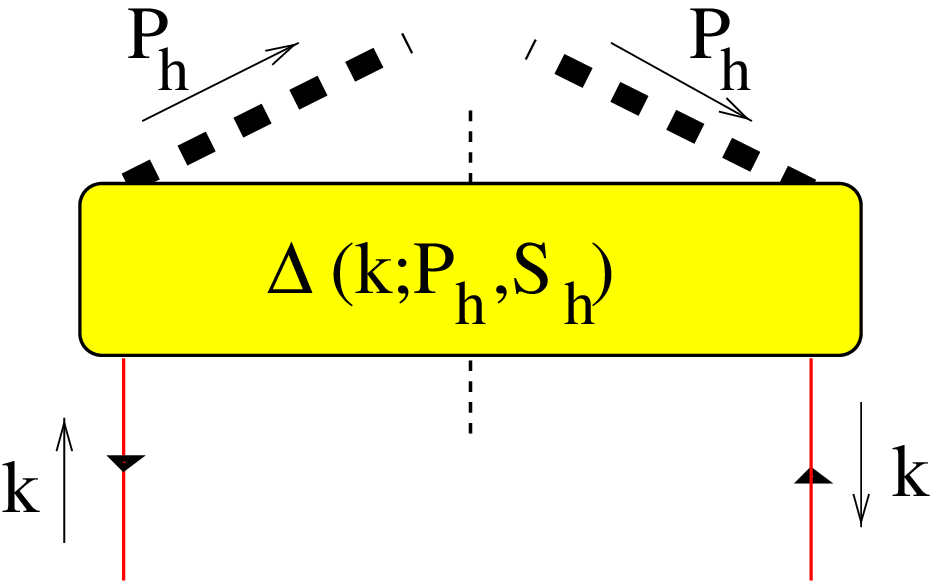}
\end{center}
\vspace{2mm}

\subsection{distribution functions}

In a particular hard process only certain Dirac projections of the quark-quark
correlation functions appear and the non-locality is restricted by the
integration over quark momenta to light-like directions.

The Dirac projections of $\Phi_{ij}(p,P,S)$ 
\be
\Phi^{{[\Gamma]}}(x) =
\left.{1\over 2}\int dp^-\,d^2\bf p_T {\rm Tr}\left(\Phi\Gamma\right)
\right|_{p^+=xP^+}
\ee
define the usual {\em distribution functions}. The functions $f_1$,
$g_1$ and $h_1$ obtained by
\bea
\Phi^{[{\gamma^+}]}(x)                    &=& {f_1(x)}\\
\Phi^{[{\gamma^+\gamma_5}]}(x)            &=& \lambda \; {g_1(x)}\\
\Phi^{[{i\sigma^{\alpha +}\gamma_5}]} (x) &=& S_T^\alpha \; {h_1(x)}
\eea
are leading in an $1/Q$ expansion, where $Q$ is the hard scale typical to the
process. $\lambda$ is the helicity of the hadron and $S_T$ the transverse part
of the spin vector. Leading functions have an
intuitive probabilistic interpretation: \\
\begin{center} \fbox{\begin{minipage}{116mm} 
\includegraphics[width=2cm]{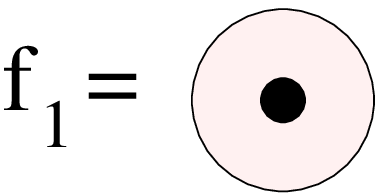}\hfill
\includegraphics[width=5cm]{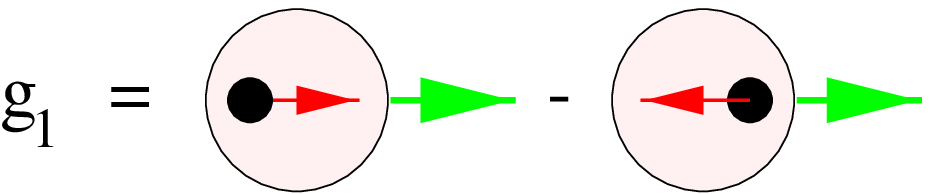}\hfill
\includegraphics[width=3.6cm]{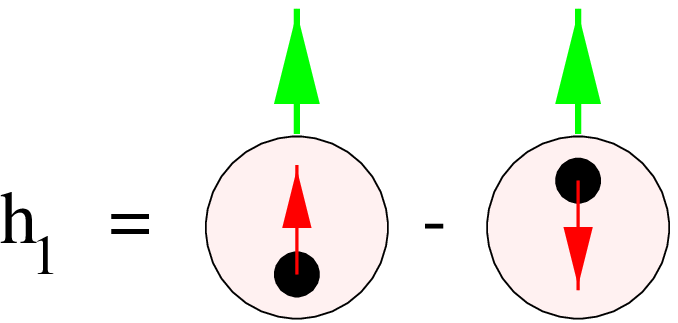}
\end{minipage}}\end{center}
\vspace{2mm}
\begin{itemize}
\item
$f_1(x)$ gives the probability of finding a quark with light-cone 
momentum fraction $x$ in the $+$ direction (and any transverse momentum).
\item
$g_1(x)$ is a chirality distribution: in a hadron that is in a positive
helicity eigenstate, it measures the probability of finding a 
right-handed quark with light-cone momentum fraction $x$ minus the
the probability of finding a left-handed quark with the same light-cone
momentum fraction. 
\item
$h_1(x)$ is a transverse
spin distribution: in a transversely polarized hadron, it measures the
probability of finding quarks
with light-cone momentum fraction $x$ polarized along the direction of
the polarization of the hadron minus the probability of finding quarks
oppositely polarized. 
\end{itemize} 
The subleading (`higher-twist') functions have no intuitive partonic
interpretation. Nevertheless, they are well defined as projections of
the quark-quark correlation functions. The pre-factor $M/P^+$ behaving like
$1/Q$ in a hard process signals the
sub-leading nature of the corresponding distribution functions
\bea
\Phi^{[{1}]}(x)                     &=& {M\over P^+}\;{e(x)}\\
\Phi^{[{\gamma^\alpha\gamma_5}]}(x) &=& {M\over P^+}\;
                                            S_T^\alpha\;{g_T(x)}\\
\Phi^{[{i\sigma^{+-}\gamma_5}]} (x) &=& {M\over P^+}\;
                                            \lambda\;{h_L(x)}.
\eea

\subsection{fragmentation functions}
Analogously, the {\em fragmentation functions} are defined by the 
Dirac projections 
\be
\Delta^{[\Gamma]}(z) = 
\left.{1\over 4z}\int dk^+\,d^2\bf k_T {\rm Tr}\left(\Delta\Gamma\right) 
\right|_{k^- = P_h^-/z}
\ee
where the leading ones are obtained by
\bea
\Delta^{[{\gamma^-}]}(z)              &=& {D_1(z)}\\
\Delta^{[{\gamma^-\gamma_5}]}(z)      &=& \lambda_h\;{G_1(z)}\\
\Delta^{[{i\sigma^{\alpha -}\gamma_5}]} (z)   
                                          &=& S_{hT}^\alpha\;{H_1(z)}
\eea
and subleading `twist' fragmentation functions by
\bea
\Delta^{[{1}]}(z)                     &=& 
{M_h\over P_h^-}\;{E(z)}\\
\Delta^{[{\gamma^\alpha\gamma_5}]}(z) &=& 
{M_h\over P_h^-}\;S_T^\alpha\;{G_T(z)}\\
\Delta^{[{i\sigma^{-+}\gamma_5}]} (z) &=& 
{M_h\over P_h^-}\;\lambda_h\;{H_L(z)}.
\eea

Note the pre-factor $M_h/P_h^-$ (scaling like
$1/Q$) accompanying the latter functions.

There are 3 additional subleading fragmentation functions: $D_T(z)$, $E_L(z)$
and $H(z)$, which are so-called ``T-odd'' functions~\cite{bjm97}. Since we
will describe 
the states of the produced hadrons by free spinors, our model leads to identical
vanishing T-odd functions. 

\section{Spectator model}

\subsection{Strategy to obtain estimates}
The purpose of our investigation is to obtain estimates for those distribution
and fragmentation functions which are experimentally poorly (or not at all)
known at present. These are, in particular, the transverse spin distribution
$h_1$, the polarized fragmentation functions
$G_1$, $H_1$ and the subleading twist functions $e(x)$, $g_T(x)$, $h_L(x)$ and
$E(z)$, $G_T(z)$, $H_L(z)$.

To this end we employ a rather simple spectator model with only a few
parameters. After fixing the parameters by phenomenological constraints we
check that the gross features of the experimentally well-known distribution 
functions $f_1(x)$ and $g_1(x)$ and the fragmentation function $D_1(z)$ are
satisfactorily reproduced.

The advantages of the model are its Lorentz covariance and its easy
extensibility to polarized and subleading twist functions. We end up with
analytic expressions for all the desired functions.

\subsection{ingredients of the model}

The ingredients of the model are exemplified for the 
distribution functions~\footnote{for more detailed information about the
model see also previous publications~\cite{jmr97,mm91,tho94,nh95}}.
\begin{itemize} 

\item
The basic idea of the spectator model is to assign a {\em definite
mass to the intermediate states} occurring in the quark-quark correlation
functions 
\begin{center}
\includegraphics[width=4cm]{phi.eps}
\hspace{10mm}
\includegraphics[width=2.9cm]{delta.eps}\\
\hspace*{4mm}$\scriptstyle (P-p)^2={M_R}^2$ \hspace{28mm} 
$\scriptstyle (k-P_h)^2={M_R}^2$
\end{center}

\item
The {\em quantum numbers} of the intermediate state are determined
by the action of the quark field operator on the hadronic state $|P,S\rangle$,
i.e. they are the quantum numbers of a {\em diquark} system
\begin{center}\fbox{\begin{tabular}{rcc}
     & \ul{spin} & \ul{isospin} \tstrut\\
     scalar diquark & {0} & {0} \tstrut\\
axialvector diquark & {1} & {1}
\end{tabular}}
\end{center}

\item
The {\em matrix element} appearing in the RHS of (\ref{eq:Phi}) is
given by 
\be
\langle X_{s} | \psi_i(0) | P, S \rangle =
\left(i \over \pslash -m\right)_{ik} \;\Upsilon^{s}_{kl} 
\ U_l(P,S)
\ee
in the case of a scalar diquark, or by
\be
\langle X_{a}^{\lambda} | \psi_i(0) | P, S \rangle =
\epsilon_\mu^{*\lambda}
\left(i \over \pslash -m\right)_{ik} \;\Upsilon^{a\mu}_{kl} 
\ U_l(P,S)
\ee
for a vector diquark. The matrix elements consist of a nucleon-quark-diquark 
vertex $\Upsilon (N)$, the Dirac spinor
for the nucleon $U_l(P,S)$, a quark propagator for the untruncated quark line
and a polarization vector $\epsilon_\mu^{*(\lambda)}$
in the case of an axial vector diquark. 

\item
For the {\em nucleon-quark-diquark vertex} we assume the following
Dirac structures \footnote{a special case of the most general
form~\cite{tho94}}:\\[2mm]  
\begin{minipage}{30mm} 
\includegraphics[width=2.6cm]{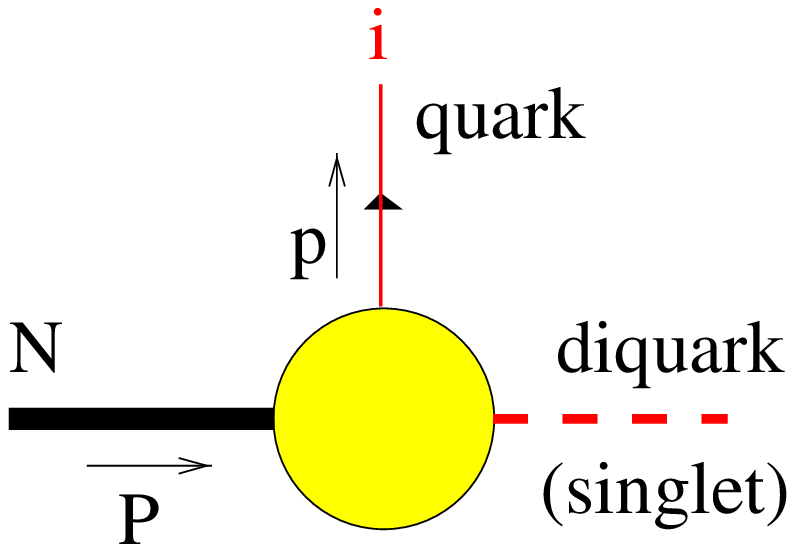}
\end{minipage} 
\begin{minipage}{79mm} 
\be \Upsilon^s (N) = {\bf 1} \ g_s(p^2)\ee
\end{minipage}\\[4mm]
\begin{minipage}{30mm}  
\includegraphics[width=2.6cm]{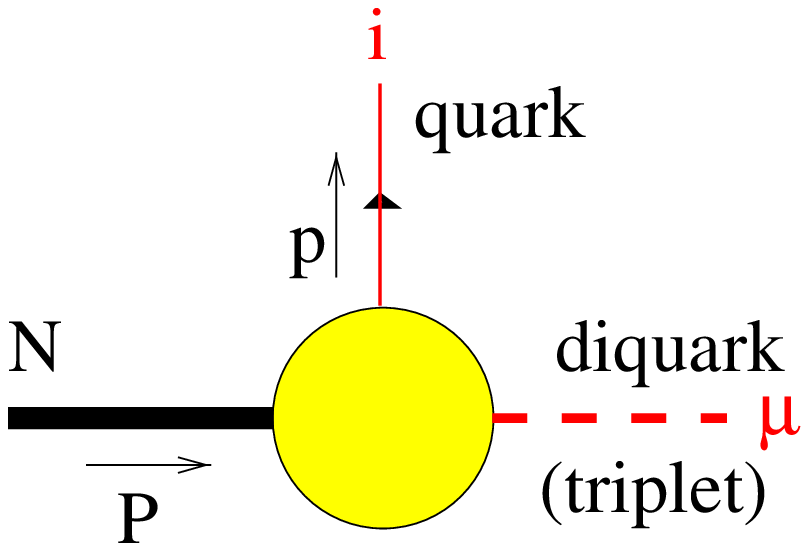}
\end{minipage} 
\begin{minipage}{79mm} 
\be \Upsilon^{a \mu} (N) = \frac{g_a(p^2)}{\sqrt{3}}\,\gamma_5\,
                        \left(\gamma^\mu+\frac{P^\mu}{M}\right) \ee
\end{minipage}

\item
The functions 
$g_{s/a} (p^2)$ are {\em form factors} that take into 
account the composite structure of the nucleon and the diquark spectator
\cite{tho94}. We use the same form factors for scalar and axial vector diquark:
\be
g_{s/a}(p^2) = N\,\frac{p^2 - m^2}{\left| p^2 - \Lambda^2 \right|^\alpha}.
\ee

\item
The {\em flavor} coupling of the proton wave function from a scalar diquark
($S_0$) and an (axial)vector diquark with isospin $0$ or $1$ ($A_0$ or $A_1$,
respectively) 
\be
|p \rangle = {1 \over \sqrt{2}}\ \vert u \ S_0\rangle +
{1 \over \sqrt{6}} \vert u \ A_0\rangle
- {1\over \sqrt{3}} \ \vert d \ A_1\rangle,
\ee
leads to the flavor relations
\bea
f_1^u &=& \frac{3}{2}\,f_1^s + \frac{1}{2}\,f_1^a \\
f_1^d &=& f_1^a
\eea
and similarly for the other functions. The coupling of the spin has already
been included in the vertices. 
\end{itemize} 
Putting all ingredients together analytic expression for the quark-quark
correlation functions are obtained
\bea
\Phi^R(p,P,S)&=&{N^2\over 2(2\pi)^3}
{\delta(p^2-2P\cdot p+M^2-M_R^2)\over|p^2-\Lambda^2|^{2\alpha}}\nonumber\\ 
&&\qquad\times
(\pslash +m)(\Pslash+M)\left(1+a_R\gamma_5\Sslash\right)(\pslash +m)
\eea
from which the distribution functions can be easily projected. The calculation
of the quark-quark correlation function for the fragmentation,
$\Delta_{ij}(k,P_h,S_h)$, proceeds along the same line.

\subsection{fixing the parameters}

The parameters of the model are fixed as follows: the power in the
denominator of the form factor, $\alpha=2$, is chosen to reproduce the 
Drell-Yan-West relation for large $x$. The mass difference $M_a-M_s$ is
motivated by the $N-\Delta$ mass difference (with group theoretical factors
properly accounted for) and the values for $\Lambda$ and $M_s$ reproduce the
experimental value for the axial charge, $g_A$.
\begin{center} 
\begin{tabular}{c|l}
parameter & fixed by: \tstrut \\ \hline\hline
{$\alpha=2$}               & Drell-Yan-West relation \tstrut \\
{$M_a-M_s=200\;{\rm MeV}$} & $N-\Delta$ mass difference \tstrut \\
{$M_s=600\;{\rm MeV}; \Lambda=0.5\;{\rm GeV}$} & $g_A=1.28$ \tstrut \\
{$N$} \tstrut & \parbox[t]{40mm}{number sum rules\\ 
                                      $\int dx f_1^u=2$; $\int dx f_1^d=1$} 
\end{tabular}
\end{center}

\section{Numerical results}
Having fixed the parameters of the model we can present the numerical results
for the {\em distribution} and {\em  fragmentation functions}. Analytical
expressions for the functions can be found in the long
write-up~\cite{jmr97}. Note that the model does 
not provide a scale dependence; but it is expected to describe physics at a low
``hadronic'' scale. Thus we compare to distributions found in the
literature at scales as low as available.

\begin{itemize} 
\item
The experimentally best known function is the distribution $f_1(x)$. We
compare our results with the parametrization from Gl\"uck, Reya and
Vogt~\cite{grv95} at the low scale $\mu_{LO}^2=0.23\;{\rm GeV}^2$. Note that the
first moments, $\int f_1(x) dx$ are exactly the same 
(normalization condition), a fact not immediately apparent from the diagram,
since we plot the combination $x*f_1(x)$. Our
distribution is narrower due to the non-inclusion of gluons and anti-quarks
(which -- if included  in our model -- would have a broadening effect). Thus we
refrain from fine-tuning parameters to obtain a better agreement with GRV ---
and are satisfied with agreement in the gross features.\\
\begin{center} \fbox{
\includegraphics[width=10.6cm]{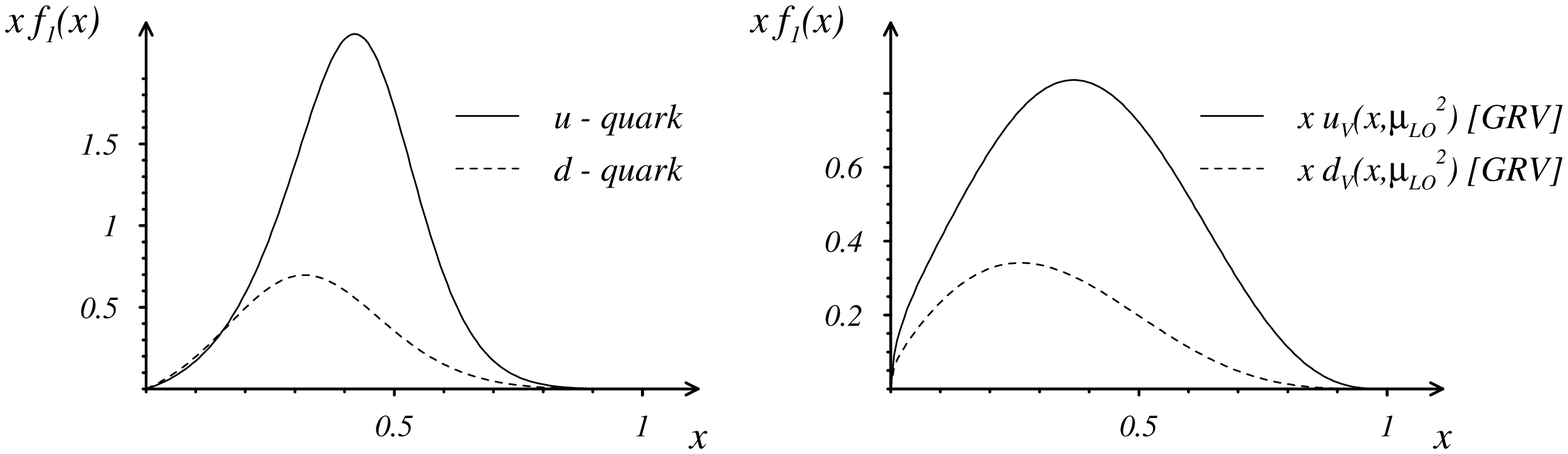}}
\end{center}

\item 
Less well experimentally determined is the helicity distribution $g_1(x)$. Our
predictions are again smaller than the distribution taken from the literature
\cite{grsv96}. Signs and positions of extrema are in reasonable agreement.
\begin{center} \fbox{
\includegraphics[width=10.6cm]{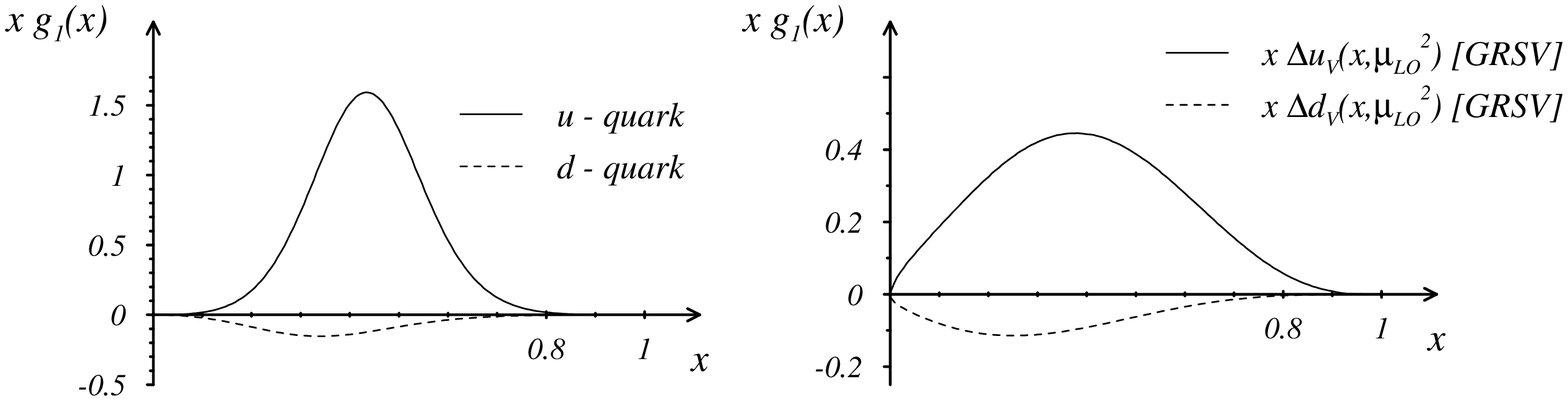}}
\end{center}

\item
Yet experimentally undetermined is the transverse spin distribution
$h_1(x)$.\\[2mm]
\begin{minipage}{60mm} 
\begin{center} \fbox{
\includegraphics[width=5.4cm]{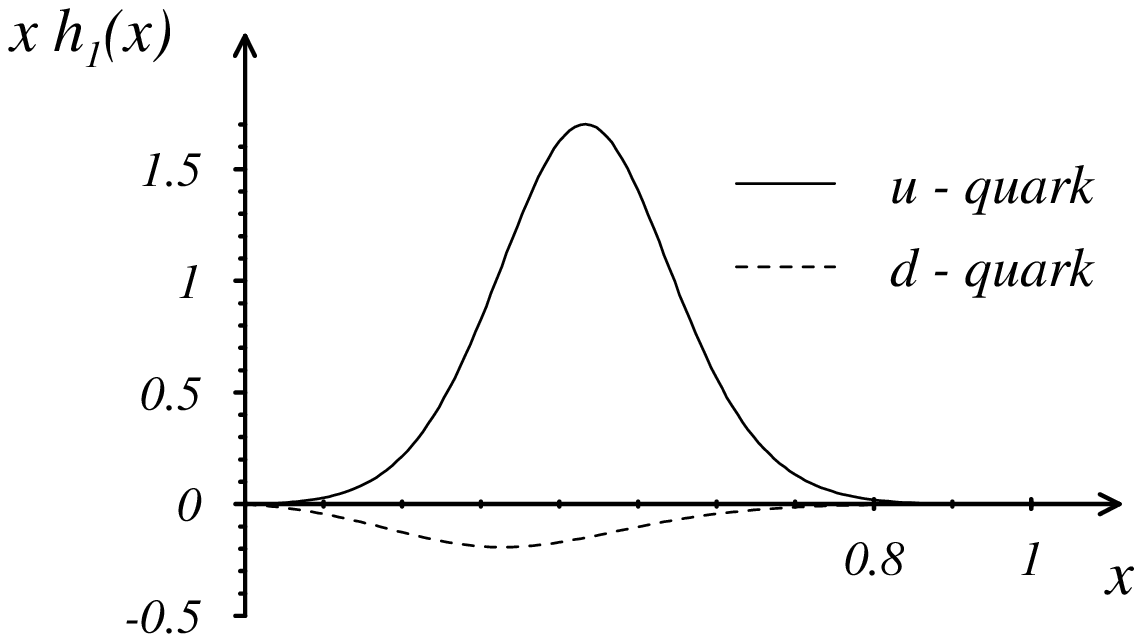}}
\end{center}
\end{minipage} 
\begin{minipage}[b]{50mm}
Within our model we obtain an estimate for this function which is numerically 
close to the result for the helicity distribution $g_1(x)$, but not identical.
\end{minipage} 
\item
As well unknown are the subleading `twist' functions like
$e(x)$ and the combination $g_2(x)=g_T(x)-g_1(x)$, for which we display our
estimates below. 
\begin{center} \fbox{
\includegraphics[width=10.6cm]{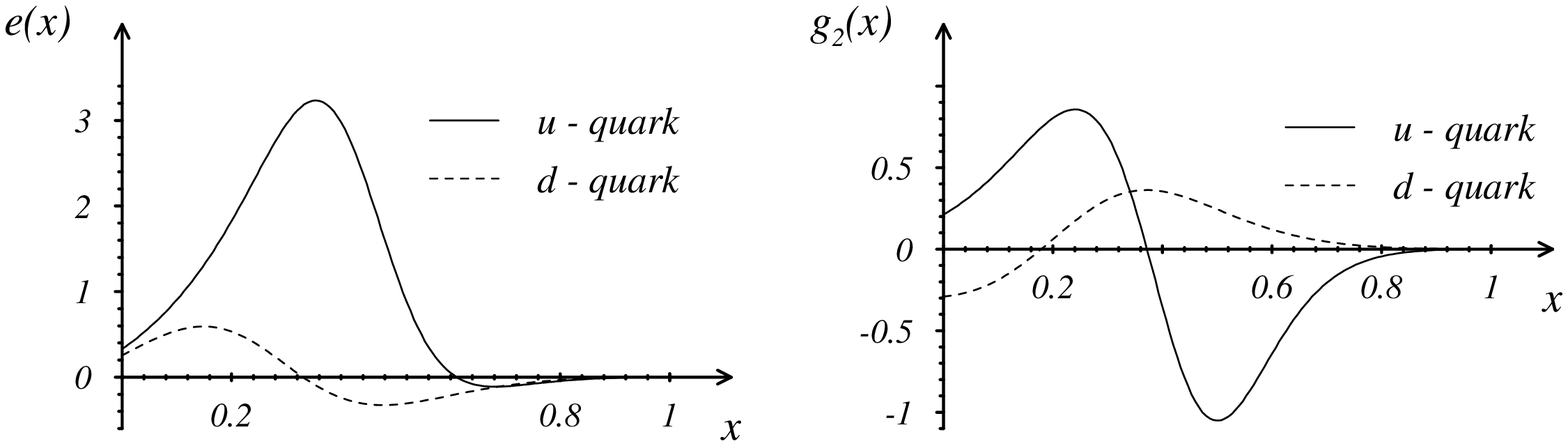}}
\end{center}

\item
On the side of the fragmentation functions only the spin-independent
information contained in $D_1(z)$ has been experimentally
investigated. Unfortunately, no parametrizations at a low ``hadronic'' scale 
are available in the literature. Thus we compare our findings with data from
the EMC collaboration~\cite{EMC} at $20\;{\rm GeV}^2$; keeping the mentioned
shortcoming 
in mind. We plot for the data the combination $D_1^{u\to p}-D_1^{\bar u\to
p}$. The latter term being predicted zero in our model (no anti-quarks
included !) this ``valence'' combination seems to be best suited for
a fair comparison. 

A second difficulty is caused by the fact that for
fragmentation functions we do not have a number sum-rule to fix the
normalization. Thus we normalize our findings such that they reproduce the
second moment, $\int z\,D_1(z) dz$ (a scale invariant quantity) as obtained
from the EMC data. 

We avoid the normalization problem for the (longitudinal and transverse)
spin-dependent fragmentation functions, $G_1(z)$ and $H_1(z)$, by showing
their ratios with $D_1(z)$. 
\begin{center} \fbox{ 
\includegraphics[width=10.6cm]{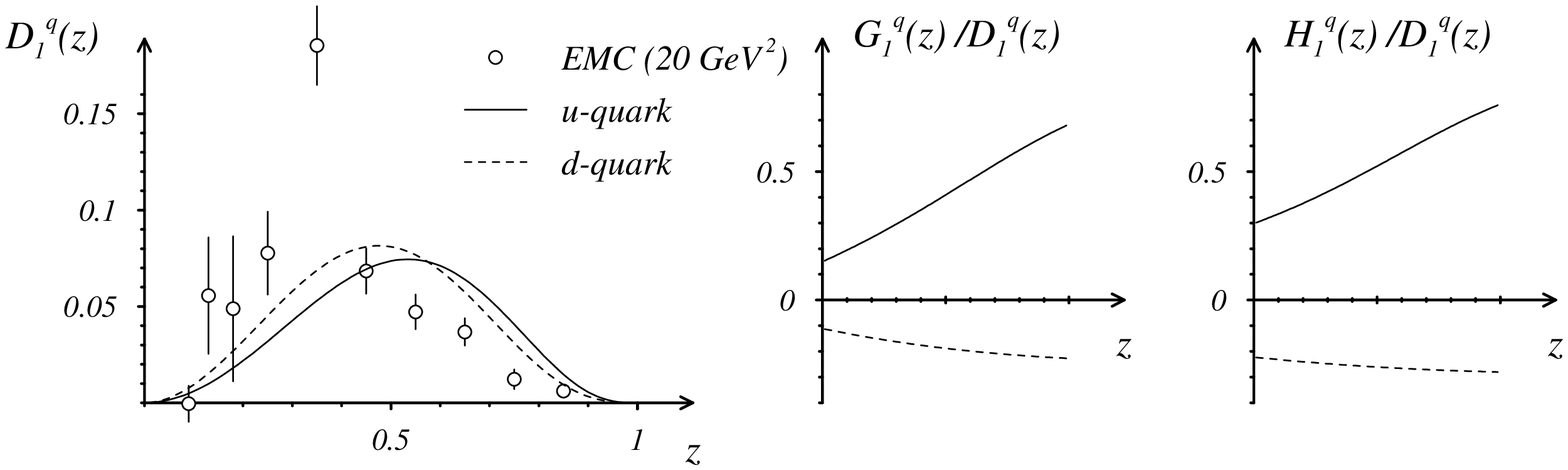}}
\end{center}

\item
The subleading `twist' fragmentation functions have an even worse chance to be
experimentally investigated in the near future than the leading spin-dependent
ones. We don't display numerical results for them, although they are
easily obtainable in the context of our model. All necessary information to
write down their analytic expressions is contained in our article~\cite{jmr97}. 

\end{itemize} 

\section{Conclusions}

We presented estimates for distribution and fragmentation functions obtained
in the context of a spectator model using representations of the functions in
terms of non-local operators.

By comparing our expressions to available parametrizations at low
``ha\-dro\-nic'' scales and to some experimental data we find 
that we can obtain reasonable qualitative agreement for $f_1(x)$, $g_1(x)$ 
and $D_1(z)$. These findings give confidence that the estimates obtained
for the ``terra incognita'' functions (transverse spin distributions,
longitudinal and transverse spin fragmentations and subleading functions)
provide a reasonable estimate of the order of magnitude of the functions and
their (large) $x$ behavior despite the simplicity of the model.

Moreover, the model predictions fulfill not only
trivial positivity constraints like $|g_1(x)|\le f_1(x)$ and $|h_1(x)|\le
f_1(x)$, but also for instance the Soffer inequality~\cite{sof95}, $2|h_1(x)|\le
(f_1(x)+g_1(x))$; the latter in our case becoming an equality \footnote{compare
the remark on a diquark model in Soffer's article~\cite{sof95}}.  

\section*{Acknowledgments}
This work is part of the scientific program of the
foundation for Fundamental Research on Matter (FOM),
the Dutch Organization for Scientific Research (NWO)
and the TMR program ERB FMRX-CT96-0008.

\end{document}